# New Online Ecology of Adversarial Aggregates: ISIS and beyond

N.F. Johnson[1], M. Zheng[1], Y. Vorobyeva[2], A. Gabriel[1], H. Qi[1], N. Velasquez[2], P. Manrique[1], D. Johnson[3], E. Restrepo[4], C. Song[1] and S. Wuchty[5,6]

[1]Dept. of Physics, University of Miami, Coral Gables, FL 33126, USA
[2]Dept. of International Studies, University of Miami, Coral Gables, FL 33126, USA
[3]Dept. of Government, Harvard University, Cambridge, MA 02138, USA
[4]Dept. of Geography and Regional Studies, University of Miami, Coral Gables, FL 33126, USA
[5]Dept. of Computer Science, University of Miami, Coral Gables, FL 33126, U.S.A.
[6]Center for Computational Science, University of Miami, Coral Gables, FL 33126, USA

**Abstract**:
Support for extremist entities – whether from the far right, or far left -- often manages to survive globally online despite significant external pressure, and may ultimately inspire violent acts by individuals having no obvious prior history of extremism. Examining longitudinal records of extremist online activity, we uncovered an ecology evolving on a daily timescale that drives online support, and we provide a mathematical theory that describes it. The ecology features self-organized aggregates (online groups such as on Facebook or another social media analog) that proliferate preceding the onset of recent real-world campaigns, and adopt novel adaptive mechanisms to enhance their survival. One of the predictions is that development of large, potentially potent online groups can be thwarted by targeting smaller ones.

Extremist entities such as ISIS (known as Islamic State) stand to benefit from the global reach and speed of the Internet for propaganda and recruiting purposes, in ways that were unthinkable for their predecessors *(1-10)*. This increased connectivity may not only facilitate the formation of real-world organized groups that subsequently carry out violent attacks (e.g. the ISIS-directed attacks in Paris, November 2015) but may also inspire self-radicalized actors with no known prior history of extremism or links to extremist leadership, to operate without actually belonging to a group (e.g. the ISIS-inspired attack in San Bernardino, December 2015) *(11)*. Recent research has used records of attacks to help elucidate group structure in past organizations for which the Internet was not a key component *(3,6,12)*, the nature of attacks by lone-wolf actors *(13)* and the relationship between general online buzz and real-world events *(14-16)*. Online buzz created by individuals that casually mention ISIS or protests is insufficient to identify any long-term build up ahead of sudden real-world events (see for example Fig. S1). This leaves open the question of how support for an entity like ISIS develops online prior to any real-world group necessarily being formed, or any real-world attack perpetrated -– whether by 'recruits' or those simply 'inspired'.

Our datasets consist of detailed second-by-second longitudinal records of online support activity for ISIS from its 2014 development onwards and, for comparison, online civil protestors across multiple countries within the past three years following the U.S. Open Source Indicator (OSI) project *(14-16)*. The online Supplementary Material (SM) provides a roadmap for the paper, data descriptions and downloads. The data shows that operational pro-ISIS and protest narratives develop through self-organized online *aggregates*, each of which is an ad hoc group of followers of an online page created through Facebook or its global equivalents such as ВКонтакте (VKontakte) at www.vk.com (Fig. 1). These generic web-based interfaces allow such aggregates to form in a language-agnostic way, and with freely chosen names that help attract followers without publicizing their members' identities. Because the focus in this paper is on the ecosystem rather than the behavior of any individual aggregate, the names are not being released. They are available on request from the authors. Pro-ISIS aggregates inhabit an online environment in which predatory entities such as police cyber-groups, individual hackers and website moderators seek to shut down pro-ISIS activity and narratives *(17,18)*. In contrast to the largely mundane chatter that may casually mention ISIS on Twitter and in aggregates focused on sport for example, pro-ISIS aggregates frequently discuss operational details such as routes for financing, technological know-how and avoiding drone strikes. We chose VKontakte for our pro-ISIS analysis as (i) pro-ISIS aggregates are shut down essentially immediately on Facebook, but not on VKontakte; (ii) it is the largest European online social networking service, with more than 350 million users; (iii) it allows multiple languages and is used worldwide; (iv) being based in Russia, it has a high concentration of users of Chechen origin focused in the Caucasus region near ISIS' main area of influence in the Levant; (v) ISIS used it to spread propaganda among the Russian-speaking population *(2)*.

Our methodology for identifying these pro-ISIS aggregates was as follows. We manually identified relevant narratives using hashtags in multiple languages, e.g. #*isn* #*khilafah* #*fisyria* #игиш (i.e. ISIS) #дауля (i.e. *dawla*, meaning 'state') #халифат (i.e. 'Caliphate'), and traced these to underlying aggregates. The specific criterion for inclusion in the list was that the group explicitly expressed its support for ISIS, publishing ISIS-related news or propaganda and/or calling for jihad in the name of ISIS. This list was fed into software Application Programming Interfaces (APIs) that expanded it by means of automated search snowballing (Fig. S2). The expanded aggregate list was then cross-checked to eliminate false identifications. New embedded links were manually searched to identify more aggregates and hashtags. We then iterated this process until closure of the aggregate list (i.e. the search led back to aggregates that were already in the list). Although labor intensive, we were able to find closure on a daily basis in real time. A similar process was followed for the civil protest data.

We uncovered 196 pro-ISIS aggregates involving 108,086 individual followers between January 1 and August 31 2015 (Fig. 1). On any given day, the total number of follows in the follower-aggregate network (i.e. total number of links that existed on that day from followers, blue nodes, into the various aggregates, red nodes, in Fig. 1 inset) ranged up to 134,857. The data provided us with bipartite graphs in which individual members belong to aggregates, but aggregates are not linked to each other except through people. This two-mode network has a highly complex temporal evolution with strong heterogeneity in both the number of follows per individual follower (i.e. the number of links emanating from a given blue node) and the number of follows per aggregate (i.e. the number of links entering a given red node, which we define as the aggregate's size); and no obvious hierarchical structure. This suggests that the follower-aggregate dynamics are driven by self-organization. Such online support is likely a necessary but not sufficient condition for any real-world actions to subsequently take place, since many additional factors can hinder real-world execution. However Fig. 2 suggests that the online proliferation of pro-ISIS or protest aggregates can indeed act as an indicator of conditions becoming right for the onset of a real-world attack campaign or mass protests respectively. We fit the trend in the creation dates of new online aggregates (Figs. 2A-B) to a well-known organizational development curve (*19*). The escalation parameter $b$ diverges at these real-world onsets (Figs. 2C-D) and follows the same mathematical dependence $(T_c - t)^{-1}$ as a wide class of physical phase transitions *(20),* with the divergence date $T_c$ matching the actual onset almost exactly (SM). The connection to physical phase transitions again suggests self-organization is a driving factor *(20)*. While such a divergence will not necessarily pre-empt attacks involving only a few individuals, such as in San Bernardino or Paris, it can help indicate an alignment of favorable conditions and has the advantage that it does not rely on *any* real-world events having yet occurred or likely dates having being circulated through social media in advance *(14-16)*. The far longer lifetimes for online aggregates of protestors in Fig. 2B as compared to pro-ISIS aggregates in Fig. 2A, makes sense as predatory online shutdown pressure was far less for the civil protestors – in particular, we found no evidence of any shutdowns in Fig. 2B, in stark contrast to Fig. 2A. Figure 2D is likely smoother than 2C for the same reason. More aggressive anti-government protests such as the sudden outburst in Venezuela in February 2014 (Fig. S6) generate an intermediate case between Fig. 2A and 2B.

We now develop a systems level theory of this online aggregate ecology. The aggregate size variations observed empirically were characterized by distinctive shark-fin shapes (Fig. 3A) with each shutdown of a pro-ISIS aggregate severing the links into that particular aggregate -- hence the abrupt drop. This fragmentation co-exists with self-organized coalescence by which individual followers sporadically link into existing aggregates while existing aggregates sporadically link into each other. Although each aggregate's precise shark-fin shape will depend on its content and noticeability to external predators, Fig. 3 shows that the system-level features are captured using only this minimal coalescence-fragmentation process. At each time-step, a phenomenological probability $\nu_{\text{coal}}$ describes the sporadic addition of 1,2,3,… etc. follows to an aggregate (coalescence of followers) whereas $\nu_{\text{frag}}$ describes the sporadic sudden shutdown of an aggregate (fragmentation of followers). Such stochastic shutdown is realistic since the predators (e.g. government monitors, individual hackers) are largely independent entities and can only shut down aggregates that they happen to find. Larger aggregates should be more noticeable, hence we can take the probability of a particular aggregate being picked for coalescence or targeted for shutdown as proportional to the aggregate's size *(21)* though this is generalizable to other algebraic forms without affecting our main findings (SM). The total number of potential follows $N$ in the system is a sum, over all potential followers, of the maximum number of aggregates that each follower is prepared to follow. The number of follows per individual can be heterogeneous, and at any time step, not all $N$ follows are necessarily used. Computer simulations of this coalescence-fragmentation process reproduce the ecology of shark-fin shapes of all sizes (Fig. 3B) with a power-law distribution $s^{-\alpha}$ for the time-average number of aggregates of size $s$ where $\alpha = 2.5$. This is similar to the empirical value of $\alpha = 2.33$ that had high goodness-of-fit ($p = 0.86$) (Fig. 3C). These shark-fin dynamics are robust in that they emerge irrespective of when we examine the model's evolution (Fig. 3B) and for any value of $N$ as a result of the model's self-similarity, i.e. the coalescence-fragmentation process generates the same dynamics across all aggregate sizes. Connecting to real-world ISIS activity, we note that the severity of ISIS attacks is also approximately power-law distributed with exponent $\alpha = 2.44$ and goodness-of-fit $p > 0.1$ *(22)*. The model can be represented mathematically by the following coupled, nonlinear differential equations describing the number $n_s$ of pro-ISIS aggregates of size $s$ ($s > 1$) over time:

$$\frac{\partial n_s(t)}{\partial t} = \frac{\nu_{\text{coal}}}{N^2}\sum_{k=1}^{s-1} k(s-k)n_k(t)n_{s-k}(t) - \frac{2\nu_{\text{coal}} s n_s(t)}{N^2}\sum_{k=1}^{\infty} k n_k(t) - \frac{\nu_{\text{frag}} s n_s(t)}{N} \qquad (1)$$

A detailed discussion of Eq. (1) is given in the SM. Like the data and computer simulation, solving Eq. (1) mathematically yields a power-law $s^{-\alpha}$ for the time-averaged aggregate size, with an exact exponent $\alpha = 2.5$ (see Refs. 23, 24 and SM for proof). The spatial independence of Eq. (1) is consistent with online interactions being largely independent of followers' separation across the globe. The first term on the right describes the formation of an aggregate of size $s$ (i.e. $s$ follows) from a smaller one through the addition of 1,2,3,.. etc. new follows; the second describes the loss of an aggregate that coalesced with another aggregate; and the third describes the fragmentation of an aggregate of size $s$. $n_{s=1}(t)$ is the pool of isolated (i.e. unused) follows at time $t$, i.e. potential 'recruits', with $\sum_{s=1}^{\infty} s n_s(t) = N$. We take $N$ to be reasonably slowly varying though this can be generalized (SM). Adding heterogeneity to the aggregate formation process (e.g. preference for similar or diverse follows) leaves the exponent $\alpha = 2.5$ unchanged, as do a variety of other generalizations (Table S2) *(23,24)*.

Our theoretical model generates various mathematically rigorous yet operationally relevant predictions. First, anti-ISIS agencies can thwart development of large aggregates that are potentially far more potent *(21)*, by breaking up smaller ones. As shown in Fig. 3D, adding a simple cost into the model for shutting down an aggregate makes this strategy actually more effective than targeting the largest aggregates (SM). Second, if anti-ISIS agencies are insufficiently active in counter-measures and hence the overall rate at which they fragment pro-ISIS clusters becomes too small, specifically if the aggregate fragmentation rate $\nu_{\text{frag}} < (N\ln N)^{-1}$, then pro-ISIS support will grow exponentially fast into one super-aggregate (Fig. S11). Third, when fragmentation rates drop below a critical value $\nu_{\text{frag}}^{\text{critical}}$, the system enters a regime in which any piece of pro-ISIS material can spread globally across the pro-ISIS support network through contagion: $\nu_{\text{frag}}^{\text{critical}} = \nu_{\text{coal}} p/q$ with $p$ and $q$ representing the probabilities of follower-to-follower transmission and follower recovery respectively (*25*). To prevent diffusion of potentially dangerous material and ideas, the fragmentation rate should be greater than $\nu_{\text{coal}} p/q$. Fourth, any online 'lone wolf' actor will only truly be alone for short periods of time (of the order of weeks in Fig. 3A for example) before being attracted into one aggregate or another through coalescence. Fifth, a systems level tool emerges for detecting the future online emergence of new ISIS-like entities, which is to employ our methodology to check if a crude power-law distribution with $\alpha$ near 2.5 begins to emerge for aggregate support surrounding a particular theme.

At a more microscopic level, the data reveals that pro-ISIS aggregates exhibit the ability to collectively adapt in a way that can extend their lifetime and increase their maximal size (Fig. 4) despite the fact that each aggregate is an ad-hoc group of followers who likely have never met, do not know each other, and do not live in the same city or country. For the civil protests, by contrast, we detected no such adaptations and no online predatory shutdowns, adding support to the notion that the pro-ISIS adaptations are a response to their high-pressured online environment. Figures 4A-C illustrate the remarkable speed, variety and novelty of these adaptations, with 15% of aggregates exhibiting name changes; 7% exhibiting flips between online visibility (i.e. content open to any VKontakte user) and invisibility (i.e. content open only to current followers of the aggregate); and 4% exhibiting reincarnation in which an aggregate disappears completely and then re-emerges at a later time with another identity but with most (e.g. >60%) of the same followers. Such reincarnation is not known to occur in real-world ecologies of living organisms. Figure 4D confirms these adaptations tend to increase not only the maximum number of followers attracted into the aggregate (maximum size) but also its lifetime. The 0.9 value for the reincarnation lifetime can be understood as follows: Reincarnation involves the aggregate temporarily disappearing, therefore an aggregate that uses reincarnation runs a high risk of losing followers since they do not know when, and with what identity, the core follower group will re-emerge. Reincarnation hence tends to be used by aggregates that are attracting unusually high predation and would otherwise have had a much shorter lifetime. Reincarnation extends this lifetime beyond its otherwise much shorter value, but not enough to reach the value of 1 corresponding to aggregates that experience less intense shutdown pressure and hence do not employ adaptations. These observations open up the possibility to add evolutionary game theoretic features into our systems-level theory in order to explain the multiple use of particular adaptations by particular aggregates, and their decision of when to adapt. A future generalized theory could prove possible employing game theoretic ideas from (*26*), for example.

More generally, our findings suggest that instead of having to analyse the online activities of many millions of individual potential actors worldwide *(27)*, interested parties can shift their focus to aggregates of which there will typically only be a few hundred. Our approach combining automated data-mining with subject-matter expert analysis and generative model building drawn from the physical and mathematical sciences, goes beyond existing approaches to mining such online data *(28-30)*. While recent reports *(31)* suggest that the amount of explicit pro-ISIS material online may have declined since summer 2015, it is possible that there is lower detection due to novel adaptations being employed -- as in Fig. 4 but now likely more sophisticated.

**Acknowledgments:** NFJ gratefully acknowledges partial support for preliminary work from Intelligence Advanced Research Projects Activity (IARPA) under grant D12PC00285 and recent funding under National Science Foundation (NSF) grant CNS1500250 and Air Force (AFOSR) grant 16RT0367. The views and conclusions contained herein are solely those of the authors and do not represent official policies or endorsements by any of the entities named in this paper.

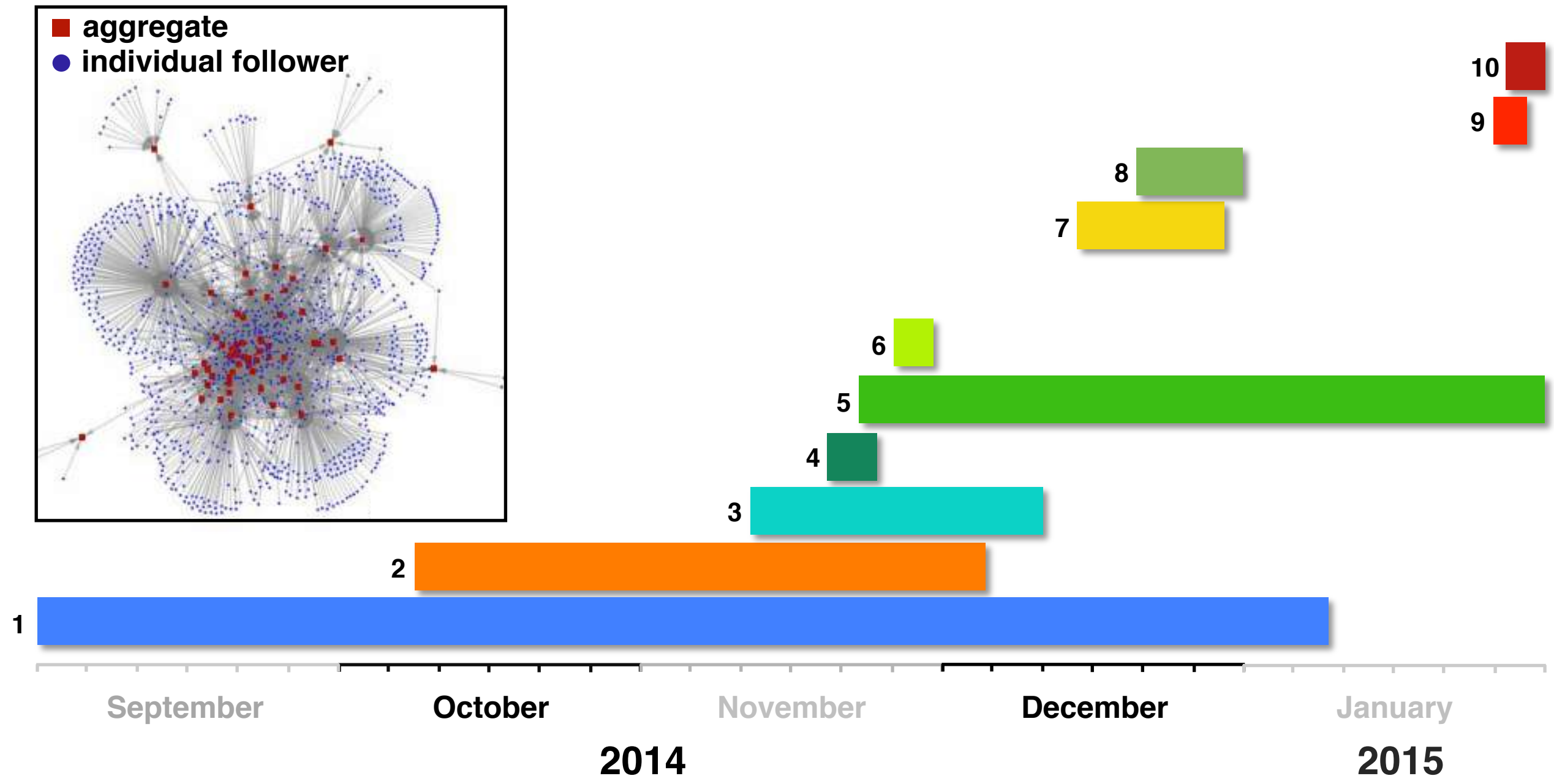

**Figure 1.** Pro-ISIS aggregates. Horizontal bars illustrate timelines of some typical pro-ISIS aggregates. Their names are available from the authors. Each timeline starts when the aggregate appears and ends when it disappears. Inset: Snapshot of part of an aggregate-follower network on January 1, 2015 showing individual followers (blue nodes) linking to pro-ISIS aggregates (red nodes). Followers can link into as many aggregates as they wish. Aggregates emerge of all sizes, where an aggregate's size is the number of follows linking into it.

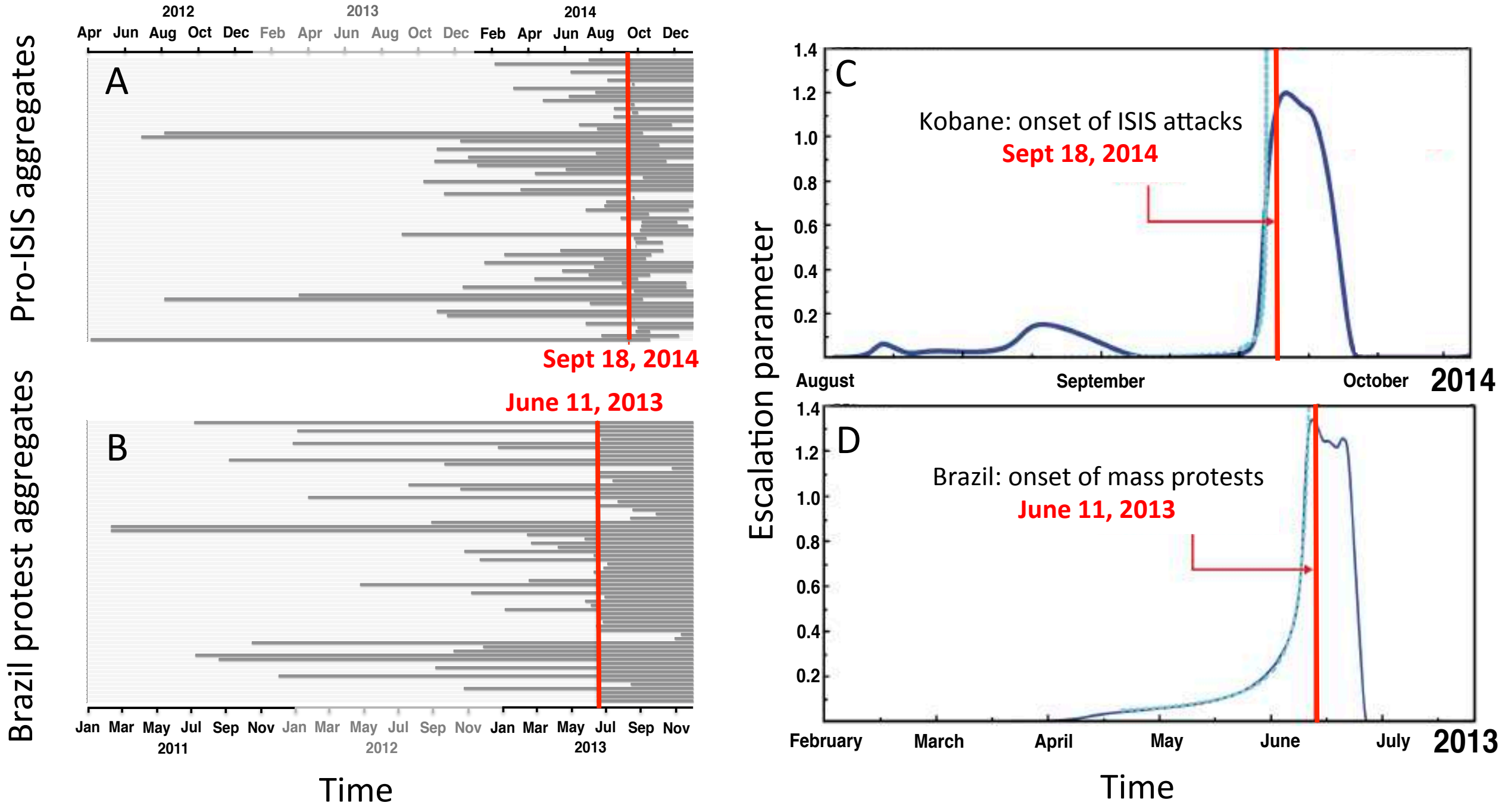

**Figure 2**: Proliferation in online aggregate creation ahead of the onsets of recent real-world campaigns (red vertical lines). **A** and **C** concern the unexpected assault by ISIS on Kobane in September 2014. **B** and **D** concern the unexpected outburst of protests in Brazil in June 2013, commonly termed the 'Brazil Winter', which involved some violence and for which we were able to collect accurate information following the IARPA OSI program *(14,15)*. Horizontal bars in **A** and **B** show timelines for (**A**) pro-ISIS aggregates on VKontakte, and (**B**) protestor aggregates on Facebook in Brazil. Each horizontal bar represents one aggregate. The aggregates are stacked separately along the vertical axis. **C** and **D**: Divergence of escalation parameter $b$ for aggregate creation (dark blue solid line) coincides with real-world onset at time $T_c$ (vertical red line). Light blue dashed line shows theoretical form $(T_c - t)^{-1}$. Subsequent decrease in both curves likely occurs for system-specific reasons associated with coalition bombings starting in **C** and loss of public interest in **D**.

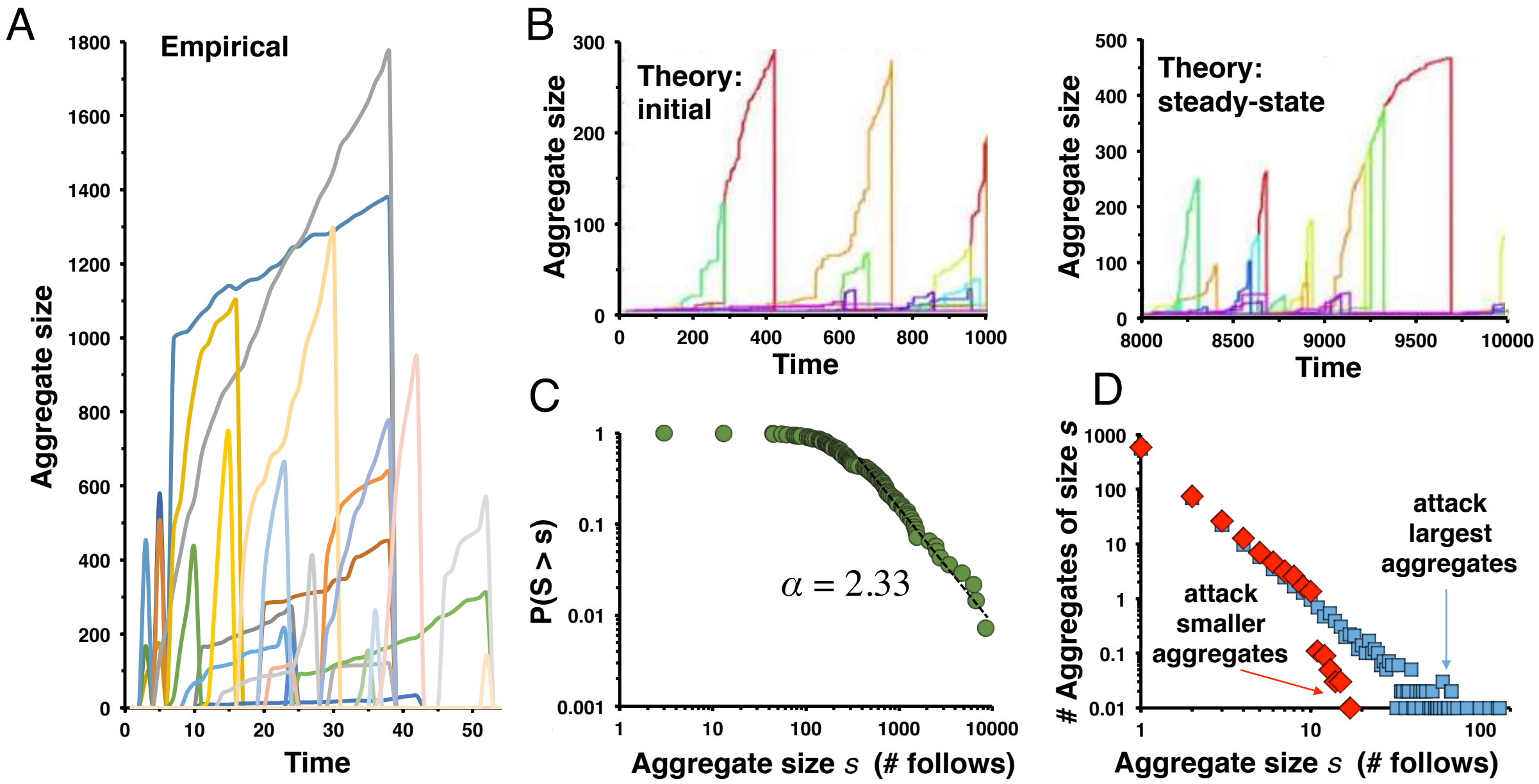

**Figure 3**: Size dynamics of pro-ISIS aggregates. **A**: Empirical size variation of pro-ISIS aggregates. Shark-fin shapes of all sizes emerge with shutdowns that are not strongly correlated. **B**: Similar results are predicted by our theoretical model, irrespective of whether we consider the model's initial (left) or steady-state (right) dynamics. Here $N = 500$, however the model's self-similar dynamics generate the same picture for any $N$ with shark-fin shapes of all sizes. An aggregate grows by an individual linking in (i.e. size increases by 1) or by an existing aggregate linking in (i.e. size increases by >1) as shown by the color change. In **B**, knowledge of the theory's microscopic dynamics allows us to denote each coalescence of a large aggregate by a color change, whereas in the empirical data **A** we maintain a constant color for each aggregate. **C**: Complementary distribution function for the observed aggregate sizes. **D**: Effect of intervention strategy involving dismantling smaller aggregates (SM). Using a larger $N$ increases the vertical and horizontal scales without changing the main results (see Fig. S10). Red diamonds: $s_{\mathrm{min}} = 10$ and $s_{\mathrm{max}} = 50$. Blue squares: $s_{\mathrm{min}} = 200$ and $s_{\mathrm{max}} = 1000$.

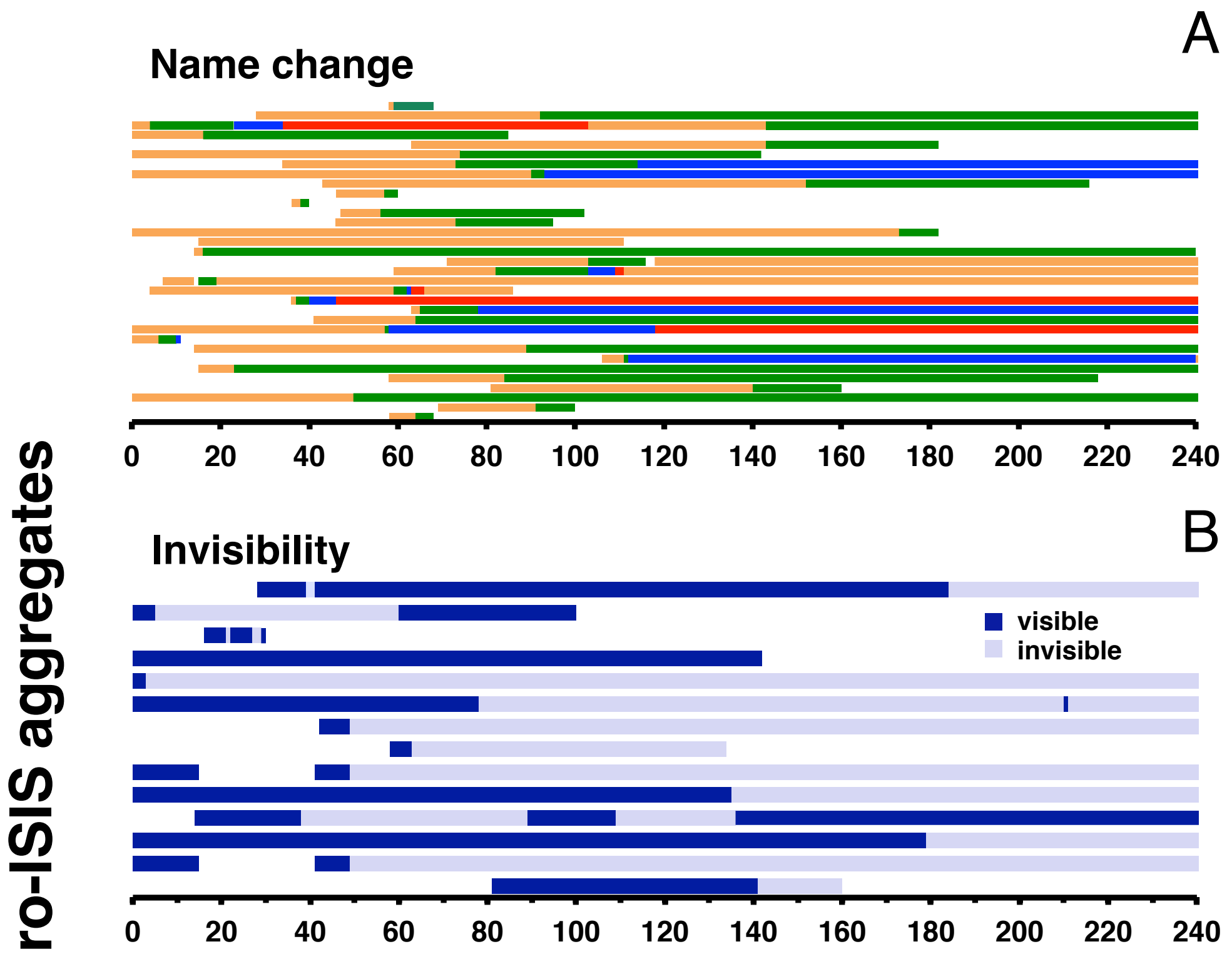

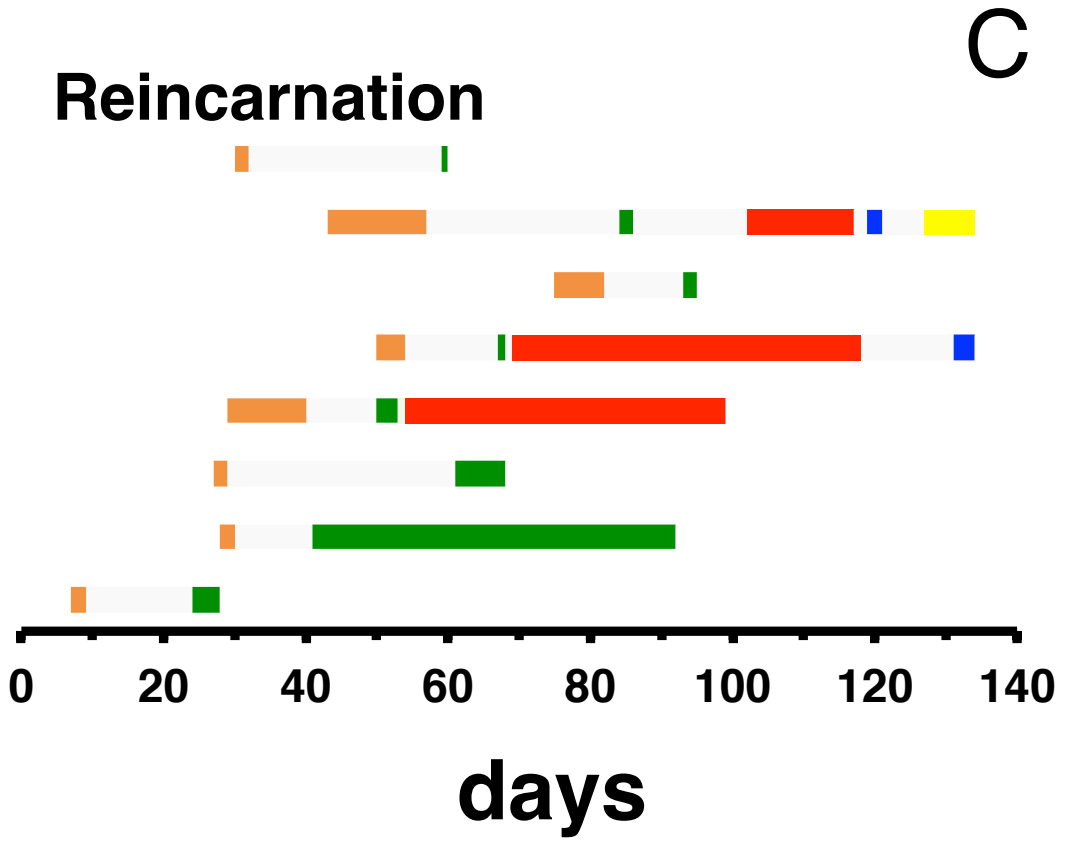

D

| Adaptation | Aggregate Rel. max. size | Aggregate Rel. life time |
|---|---|---|
| None | 1 | 1 |
| All | 1.4 | 1.4 |
| Name change | 1.1 | 1.5 |
| Invisibility | 1.5 | 2.4 |
| Reincarnation | 2.0 | (0.9) |

**Figure 4**: Evolutionary adaptations. **A-C**: Sample of pro-ISIS aggregate timelines showing evolutionary adaptations (shown by switches in colors) that tend to increase an aggregate's maximum attained size and extend its lifetime (**D**). Time measured in days from January 1, 2015. In **A**, the switch in colors within a given timeline indicates a switch in aggregate name. **B**: dark blue means aggregate is visible (i.e. content open to any VKontakte user) while light blue means invisible (i.e. content open only to current followers of the aggregate). **C**: aggregate has a specific initial identity (orange), then disappears from the Internet for an extended time (white), then reappears with another identity shown by switch in color. **D**: relative maximum aggregate size and relative lifetime for particular adaptations and their combinations, given as average values relative to the values for aggregates employing no adaptation. 'All' corresponds to aggregates that use name-change, invisibility and reincarnation. See text for explanation of (0.9) entry.